\documentclass[superscriptaddress,twocolumn]{revtex4-1}%
\usepackage[ansinew]{inputenc}
\usepackage{graphicx}
\usepackage{tikz}
\usepackage{amsmath}
\usepackage{bm}
\usepackage{layout}
\usepackage{float}
\usepackage{txfonts}
\usepackage{amsfonts}
\usepackage{amssymb}
\usepackage{array}%
\usetikzlibrary{decorations.pathreplacing}
\newcolumntype{L}[1]{>{\raggedright\let\newline\\\arraybackslash\hspace{0pt}}m{#1}}

\begin{document}

\title{Requirements for a loophole-free photonic Bell test using imperfect setting generators}

\begin{abstract}
Experimental violations of Bell inequalities are in general vulnerable to
so-called “loopholes.” In this work, we analyse the characteristics of a
loophole-free Bell test with photons, closing simultaneously the locality,
freedom-of-choice, fair-sampling (i.e.\ detection), coincidence-time, and
memory loopholes. We pay special attention to the effect of excess
predictability in the setting choices due to non-ideal random number
generators. We discuss necessary adaptations of the CH/Eberhard inequality
when using such imperfect devices and -- using Hoeffding's inequality and
Doob's optional stopping theorem -- the statistical analysis in such Bell tests.

\end{abstract}
\date{\today}%

\author{Johannes Kofler}%
%

\affiliation
{Max Planck Institute of Quantum Optics (MPQ), Hans-Kopfermann-Straße 1, 85748 Garching/Munich, Germany}%
%

\author{Marissa Giustina}%
%

\affiliation
{Institute for Quantum Optics and Quantum Information (IQOQI), Austrian Academy of Sciences, Boltzmanngasse 3, 1090 Vienna, Austria}%
%

\affiliation
{Quantum Optics, Quantum Nanophysics, and Quantum Information, Faculty of Physics, University of Vienna, Boltzmanngasse 5, 1090 Vienna, Austria}%
%

\author{Jan-{\AA}ke Larsson}%
%

\affiliation{Institutionen for Systemteknik, Link\"{o}%
pings Universitet, SE-58183 Link\"{o}ping, Sweden}%
%

\author{Morgan W. Mitchell}%
%

\affiliation
{ICFO -- Institut de Ciencies Fotoniques, The Barcelona Institute of Science and Technology, 08860 Castelldefels (Barcelona), Spain}%
%

\affiliation{ICREA -- Instituci\'{o}
Catalana de Recerca i Estudis Avan\c{c}ats, 08015 Barcelona, Spain}%
%

\maketitle

\section{Introduction}

Bell's theorem~\cite{Bell1964} about the incompatibility of a local realist
world view with quantum mechanics is one of the most profound discoveries in
the foundations of physics. Since the first experimental quantum violation of
Bell's inequality~\cite{Free1972}, countless experimental tests have been
performed with various different physical systems, closing all major
\textquotedblleft loopholes". While it is unlikely that nature exploits these
loopholes, let alone different ones for different experiments, there are at
least two reasons why a loophole-free test is of great relevance: Firstly, a
definitive ruling on local realism is of central importance to our
understanding of the physical world. Secondly, there are quantum information
protocols whose security is based on Bell's inequality, and eavesdroppers
could actively exploit the loopholes.

This work is structured as follows: We first briefly review Bell's derivation
and the five major loopholes -- the locality, freedom-of-choice, fair-sampling
(detection), coincidence-time, and memory loopholes (section II). Then, we
give an analysis of how a photonic Bell test can simultaneously close all of
them. This involves a discussion of the CH/Eberhard inequality (section III),
whose low detection efficiency requirement is essential given the current
status of equipment and technology. We outline the necessary space-time
arrangement (section IV) and show how to take into account -- by adapting the
CH/Eberhard inequality -- imperfect random number generators that sometimes
choose settings outside the allowed space-time interval or are for some other
reason partially predictable beyond the \textit{a priori} probability (section
V). Finally, while allowing both bias and excess predictability of the
settings, we demonstrate how to apply Hoeffing's inequality and Doob's
optional stopping theorem to achieve high statistical significance of a Bell
inequality violation within feasible experimental run-time (section VI).
Readers who are familiar with loopholes in Bell tests and the CH/Eberhard
inequality can skip to section IV.

\section{Bell's theorem and loopholes}

Let us consider the simplest scenario of only two parties called Alice and
Bob, who perform measurements on distant physical systems. Alice's and Bob's
measurement settings are labeled with $a$ and $b$, and their outcomes are
denoted by $A$ and $B$ respectively. There are essentially two versions of
Bell's theorem:

\textit{Deterministic local hidden variable models.} \textit{Determinism}
states that hidden variables determine the outcomes, which are then functions
of the form $A=A(a,b,\lambda)$, $B=B(a,b,\lambda)$. \textit{Locality} demands
that the local outcomes do not depend on the distant setting:%
\begin{equation}
A=A(a,\lambda),\;B=B(b,\lambda).
\end{equation}
The original 1964 version of Bell's theorem~\cite{Bell1964} is based on the
assumptions of \textit{perfect anticorrelation} and \textit{locality} which
imply determinism. The assumption of perfect anticorrelation was later avoided
by Clauser, Horne, Shimony, and Holt (CHSH) in the derivation of their famous
inequality~\cite{Clau1969}.

\textit{Stochastic local hidden variable models.} Following
Refs.~\cite{Bell1971,Clau1974}, in the 1976 version of Bell's
theorem~\cite{Bell1976} the assumptions are relaxed to include stochastic
models. There, hidden variables only define probabilities for the outcomes,
$P(A|a,b,B,\lambda)$, $P(B|a,b,A,\lambda)$, and a joint assumption called
\textit{local causality} (or \textit{Bell locality}) demands that the joint
probability of Alice's and Bob's outcomes factorizes as follows:%
\begin{equation}
P(A,B|a,b,\lambda)=P(A|a,\lambda)\,P(B|b,\lambda).
\end{equation}
This is equivalent to assuming \textit{outcome independence}
$P(A|a,b,B,\lambda)=P(A|a,b,\lambda)$ as well as \textit{setting independence}
(or \textit{parameter independence}) $P(A|a,b,\lambda)=P(A|a,\lambda)$, with
similar expressions for Bob's outcome probability~\cite{Jarr1984}.

The world view in which all physical phenomena can be described by local
hidden variables is often referred to as \textit{local realism}. While local
causality is implied by the conjunction of determinism and locality, the
opposite implication is not true. Nonetheless, the two classes of local hidden
variable models are mathematically equivalent in the sense that deterministic
models are special cases of stochastic ones (where all probabilities are 0 or
1), and that every stochastic model can be viewed as a mixture of
deterministic ones~\cite{Fine 1982,Hall2009}. Physically, however, the
difference is significant. It is conceivable to adhere to a stochastic world
view in which the hidden variables only define probabilities, rejecting a
hidden determinism, although this determinism might mathematically exist and
explain the probabilities.

In addition to local causality (or, stronger, determinism and locality) there
is another essential assumption in the derivation of every Bell inequality
called \textit{freedom of choice} (or \textit{measurement independence}). It
demands that the distribution $\rho$ of the hidden variables $\lambda$ is
statistically independent of the setting values:%
\begin{equation}
\rho(\lambda|a,b)=\rho(\lambda). \label{eq FoC}%
\end{equation}
By Bayes' theorem, this assumption can also be written as $\rho(a,b|\lambda
)=\rho(a,b)$. The freedom-of-choice assumption was first pointed out in a
footnote in Ref.\ \cite{Clau1974} and later discussed in an exchange
\cite{Bell1976,Shim1976,Bell1977}, which is reprinted in~\cite{Bell1985}.

Bell's theorem states that the joint assumption of local hidden variables and
freedom of choice enables the derivation of inequalities that put local
realist bounds on combinations of probabilities for Alice's and Bob's
measurement results. In Bell experiments, measurements on entangled quantum
states can violate Bell's inequality and thus refute the existence of local
hidden variables.

The translation from any mathematical expression to a physical experiment
employs further physical assumptions, which may render an experimental Bell
violation vulnerable to a local realist explanation. In the following, we
discuss the five main \textquotedblleft loopholes" in Bell tests. For further
details on the assumptions in Bell's theorem, the use of entanglement in Bell
experiments, and the loopholes that can arise, we refer the reader to the
recent reviews~\cite{Pan2012,Brun2014,Wise2014,Lars2014}.%
\begin{table*}[t]%
\begin{tabular}{| L{3.0cm} | L{3.0cm} | L{3.0cm} | L{8.0cm} |}
\hline Minimal assumptions & Auxiliary assumptions & Loopholes & Closed by \ldots \\
\hline\hline Outcome and setting independence &   & Locality
loophole & space-like separation between the outcome events and between each outcome and the distant setting choice event \\
\hline Freedom of choice &   & Freedom-of-choice loophole &
(for photonic experiments:)\ space-like separation between each pair emission event and the setting choice events \\
\hline   & Fair sampling & Fair-sampling (detection) loophole &
violation of an inequality free of the fair-sampling assumption (e.g.\ CH/Eberhard) or explicit demonstration of sufficiently large detection efficiency (e.g.\ for CHSH) \\
\hline   & Fair coincidences & Coincidence-time loophole &
using fixed time slots or (for CH/Eberhard) a window-sum method for identifying coincidences \\
\hline   & No memory & Memory loophole & sufficiently many measurement trials, no i.i.d. assumption \\
\hline
\end{tabular}\caption{Summary of the five main loopholes in Bell experiments. The
assumptions of outcome and setting independence as well as freedom
of choice are minimal in the sense that they enter the derivation of
any Bell inequality. The corresponding loopholes are closed by the
spatio-temporal construction of the experiment and the means of
choosing settings. The other three loopholes are related to
auxiliary assumptions and are closed by a suitable choice of Bell
inequality (or additional tests) as well as appropriate data
analysis. See main text for
further details and references.}\label{table}%
\end{table*}%

\subsection{The locality loophole}

The locality loophole refers to the possibility of violating outcome or
setting independence via subluminal or luminal influences between the two
outcomes or from one setting to the distant outcome. It is generally
acknowledged that the best possible way to close the loophole is to invoke
special relativity. Space-like separating the two outcome events enforces
outcome independence, and space-like separating each party's independent
setting choice event from the opposite party's outcome event enforces setting
independence. In this way, the locality loophole is considered to have been
closed for photons by the experiments
\cite{Weih1998,Sche2010,Erve2014,Gius2015,Shal2015}, and with NV centers by
the experiment\ \cite{Hens2015}.

This, however, rests on the assumption that there were no prior influences for
the setting choice events that could have been communicated to the distant
party. Deterministic setting mechanisms as, e.g., the periodic switching used
in \cite{Aspe1982}, are predictable into the future and thus in principle
still allow a local realist explanation \cite{Zeil1986} unless restrictions
are imposed on the information communicated.

\subsection{The freedom-of-choice loophole}

The freedom-of-choice loophole refers to the possibility that the
freedom-of-choice condition $\rho(\lambda|a,b)=\rho(\lambda)$ fails due to an
influence of the hidden variables on the setting choices, or an influence of
the setting choices on the hidden variables, or more generally due to a common
influence on both the setting choices and the hidden variables.

As with the locality loophole, space-like separation allows an experiment to
exclude certain influences within any local theory. For example, space-like
separation of the pair generation from the setting choices eliminates the pair
generation as a possible influence. This has been achieved in the
experiments\ \cite{Sche2010,Erve2014,Gius2015,Shal2015}. However, again it is
not possible to exclude all possible influences in this way, because these
could in principle extend arbitrarily far into the past.

Note that freedom of choice does not require the factorization $\rho
(a,b)=\rho(a)\,\rho(b)$. However, if the setting choices are not space-like
separated with respect to each other, then one of the outcome events will
always be in the future light cone of the distant setting event, leaving the
locality loophole open.

A second, complementary way to address the freedom-of-choice loophole is to
derive the setting choices from events that are plausibly beyond the control
of hidden variables, for example spontaneous emission, chaotic evolution,
human decision-making, or cosmic sources. A Bell inequality violation using
one or more of these sources can exclude local realist theories in which the
setting events are unpredictable, pushing the unexcluded theories in the
direction of a full determinism (c.f.\ Sec.\ II.F).

\subsection{The fair-sampling (detection) loophole}

The fair-sampling assumption states that the ensemble detected by Alice and
Bob is representative of the total emitted ensemble. This is the case if the
detection efficiency depends only on the hidden variable and not on the local
setting. Unfair sampling opens the fair-sampling (or detection)
loophole\ \cite{Pear1970}.

Inequalities that make use of the fair-sampling assumption in their
derivation, such as the CHSH inequality \cite{Clau1969}, can be rendered
immune to the fair-sampling loophole only by explicitly demonstrating
sufficiently large detection efficiency or by incorporating the undetected
events into the inequality \cite{Bell1971}. This latter, more elegant approach
-- not assuming fair-sampling in the first place -- is used in the derivation
of the Clauser-Horne (CH) \cite{Clau1974} and the Eberhard inequalities
\cite{Eber1993}. The fair-sampling loophole has been closed for
atoms\ \cite{Rowe2001,Mats2008,Hofm2012}, superconducting
qubits\ \cite{Ansm2009}, and NV centers\ \cite{Hens2015}. Using
superconducting detectors, it has also been closed for
photons\ \cite{Gius2013,Chri2013,Gius2015,Shal2015}.

\subsection{The coincidence-time loophole}

The fair-coincidence assumption states that the statistics of the identified
pairs are sufficiently representative of the statistics of all detected pairs,
had they been correctly identified. In experiments where (near-)coincident
arrival times are used to identify which detections belong to a pair, the
assumption is fulfilled if the local detection time depends only on the hidden
variable and not on the local setting. Unfair coincidences open the
coincidence-time loophole\ \cite{Lars2004}.

This loophole arises in any situation where a (setting-dependent) shift in
detection time could alter the number of identified pairs; it is especially
applicable to continuous-wave photonic experiments. The loophole can be closed
using locally predefined time-slots or (for the CH/Eberhard inequality) by
employing a window-sum method for coincidence-based identification of
pairs\ \cite{Lars2013,Knil2015}. Regarding photonic experiments, the loophole
was closed in\ \cite{Ague2012,Chri2013,Gius2013,Lars2013,Gius2015,Shal2015}.

\subsection{The memory loophole}

One can imagine a situation in which the experimental apparatuses use memory
of the previous measurements to skew the apparent significance of a violation.
In this case, say, the probability for Alice to find outcome $A^{(m)}$ in the
$m$-th measurement can depend not only on her current setting $a^{(m)}$ and
hidden variable $\lambda$, but also on the $m-1$ previous settings and
outcomes on her side ($a^{(1)},...,a^{(m-1)},A^{(1)},...,A^{(m-1)}$, one-sided
memory) and maybe also on Bob's side ($b^{(1)},...,b^{(m-1)},B^{(1)}%
,...,B^{(m-1)}$, two-sided memory), and vice versa for Bob's outcome
probability for $B^{(m)}$ \cite{Barr2002,Acca2000,Gill2003,Gill2003b}. Then,
the no-memory assumption that successive measurement trials are
i.i.d.\ (independent and identically distributed) is not valid.

The memory loophole does not change a Bell inequality's local realist bound
but forbids quantifying the statistical significance of a Bell test by the
amount of conventional standard deviations between the observed Bell value and
the local realist bound. The loophole could in principle be closed by using
separate apparatuses and space-like separation of each of Alice's measurements
from all of Bob's measurements. However, this is technologically unfeasible.
Thus, a more useful approach is to apply statistical methods, such as
hypothesis testing, that can -- without the assumption of i.i.d.\ measurement
trials -- bound the probability that the data can be explained by a random
variation of a local hidden variable model.

Table I summarizes the assumptions used in derivations of Bell inequalities as
well as the corresponding loopholes and the procedures for closing them.

\subsection{Additional assumptions and unclosable loopholes}

By attributing significance to space-like separation, one implicitly assumes
that one can localize key events to particular space-time regions. For
example, space-like separation of the setting choices from the detection
events closes the locality loophole, but requires that the setting choices are
independent of prior conditions. This break between the past and the present
means that closure of the locality loophole can only be attempted within
non-deterministic (i.e.\ stochastic) local realism. Within determinism, the
settings would also be deterministic and thus predictable arbitrarily far in
the past, rendering space-like separation impossible. Similarly, using
space-like separation to close the freedom-of-choice loophole can only
eliminate theories in which the hidden variable is created in a defined
space-time region (e.g.\ at the down-conversion event in a photonic experiment).

Likewise, arguments based on space-like separation of the detection events
from the distant setting choices requires that one knows when the measurement
is complete. In all practical scenarios for Bell tests, there is an
identifiable time window in which a microscopic observable such as the
polarization of a single photon becomes amplified into a macroscopic
observable such as a large number of electrons moving in a wire. Usually this
conversion to a macroscopic event is taken as the time of the measurement, but
there is no logical contradiction in assuming that the measurement happens
later (\textquotedblleft collapse locality loophole"\ \cite{Kent2005}).

The general feature of all these arguments is that a loophole-free Bell test
is possible only when a set of reasonable assumptions about the physical
working of the experimental setup is made. Experiments can shift hypothetical
effects to more and more absurd scales but can never fully rule them out. In
particular, it is in principle impossible to rule out \textquotedblleft
superdeterminism" \cite{Bell2004}, a world constructed such that equation
(\ref{eq FoC}) cannot be fulfilled. Therefore, strictly speaking the locality
and freedom-of-choice loopholes can only be addressed (i.e.\ closed within
some assumptions) and cannot be closed in general.

Finally, every Bell test needs to rest on metaprinciples, most notably that
the classical rules of logic hold. In 2015, three different groups were able
to perform \textquotedblleft loophole-free" Bell
tests\ \cite{Hens2015,Gius2015,Shal2015}.

\section{The CH/Eberhard inequality}

Eberhard's derivation\ \cite{Eber1993} considered a source that produces
photon pairs where the polarization of one photon of every pair is measured by
Alice with setting $a_{1}$ or $a_{2}$, while the other photon's polarization
is measured by Bob with setting $b_{1}$ or $b_{2}$. We label the outcome or
\textquotedblleft fate" (given by the hidden variable) of every photon by
`$+$', `$-$', or `$0$', which denotes being detected in the first
(\textquotedblleft ordinary") output beam of the polarizer, being detected in
the second (\textquotedblleft extraordinary") beam, or remaining undetected,
respectively. We denote joint fates for outcomes $A$ (for Alice) and $B$ (for
Bob) by $AB$ with $A,B\in\{+,-,0\}$.

Eberhard considered $N^{\prime}$ pairs emitted for each of the four setting
combinations $a_{i}b_{j}$ with $i,j\in\{1,2\}$. For setting combination
$a_{i}b_{j}$ we denote the number of joint outcomes $A$ and $B$ by
$n_{AB}(a_{i}b_{j})$. Note that pairs with joint fate $00$ also count as
pairs. Hence, $%
{\textstyle\sum\nolimits_{A,B\in\{+,-,0\}}}
n_{AB}(a_{i}b_{j})=N^{\prime}$ for each setting combination $a_{i}b_{j}$.

In hidden variable theories, the results for mutually exclusive measurements
exist simultaneously. Locality demands that the local fate of a photon must
not depend on the distant measurement setting. Freedom of choice assumes that
the experimenters' settings are independent of the designated fate. Under
these assumptions, Eberhard's inequality bounds the expectation value of a
certain combination of outcome numbers \cite{Eber1993}:%
\begin{align}
&  \left\langle \!\right.  +n_{++}(a_{1}b_{1})-n_{+-}(a_{1}b_{2})-n_{+0}%
(a_{1}b_{2})\nonumber\\
&  \ \ -n_{-+}(a_{2}b_{1})-n_{0+}(a_{2}b_{1})-n_{++}(a_{2}b_{2})\left.
\!\right\rangle \leq0. \label{eq J}%
\end{align}

The logical bound of the inequality is $N^{\prime}$, which can be attained by
a model (violating local realism and/or freedom of choice) where all
$N^{\prime}$ pairs for settings $a_{1}b_{1}$ lead to outcome $++$ and no pairs
in the other setting combinations ever contribute to the five positive terms.
The quantum bound is $(\!\sqrt{2}\!-\!1)\,N^{\prime}/2\approx0.207\,N^{\prime
}$, which can be attained for perfect detection efficiency (i.e.\ absence of
outcomes $0$) on both sides and maximally entangled states. However, for
imperfect detection efficiency (i.e.\ occurrence of outcomes $0$),
non-maximally entangled states achieve better violation.

Until now, the derivation has assumed that there was the same number of pairs
($N^{\prime}$) in each of the four setting combinations. Experiments are not
likely to obey this strict constraint, but rather to produce a different
number of pairs for every combination. In general, this invalidates the
Eberhard inequality (\ref{eq J}), as can be seen by considering the case where
the setting $a_{1}b_{1}$ is used more often than the others, which will
increase the $n_{++}(a_{1}b_{1})$ contribution (see
Refs.\ \cite{Chri2013,Kofl2013}). A solution is to introduce conditional
probabilities $p_{AB}(a_{i}b_{j})$ for outcomes $AB$ given settings
$a_{i}b_{j}$. As the original Eberhard inequality holds when an equal number
of trials is measured in each setting combination, and since under freedom of
choice every setting is chosen independently from the source, the same form of
inequality holds for the conditional probabilities:%
\begin{align}
&  +p_{++}(a_{1}b_{1})-p_{+-}(a_{1}b_{2})-p_{+0}(a_{1}b_{2})\nonumber\\
&  -p_{-+}(a_{2}b_{1})-p_{0+}(a_{2}b_{1})-p_{++}(a_{2}b_{2})\leq0.
\label{eq J norm}%
\end{align}
The logical bound of this inequality is $1$, and the quantum bound is
$(\!\sqrt{2}\!-\!1)/2\approx0.207$. One may drop the distinction between
outcomes `$-$' and `$0$' in the Eberhard inequality (\ref{eq J}). Blocking the
extraordinary beam such that all `$-$' events become `$0$' events, the
normalized Eberhard inequality (\ref{eq J norm}) is reduced to a
one-detector-per-side form with coincidences and exclusive singles
(i.e.\ detections on exactly one side):%
\begin{equation}
J\equiv p_{++}(a_{1}b_{1})-p_{+0}(a_{1}b_{2})-p_{0+}(a_{2}b_{1})-p_{++}%
(a_{2}b_{2})\leq0. \label{eq J new}%
\end{equation}
We can define the probabilities of \textit{singles} (photon detections in one
particular output beam regardless of the outcome on the other side):%
\begin{align}
p_{+}^{\text{A}}(a_{1})_{b_{2}}  &  \equiv p_{++}(a_{1}b_{2})+p_{+0}%
(a_{1}b_{2}),\label{eq sA}\\
p_{+}^{\text{B}}(b_{1})_{a_{2}}  &  \equiv p_{++}(a_{2}b_{1})+p_{0+}%
(a_{2}b_{1}), \label{eq sB}%
\end{align}
Here, the singles probabilities were defined for a particular distant setting,
namely $b_{2}$ and $a_{2}$, respectively. However, due to locality,
no-signaling must be fulfilled:%
\begin{align}
p_{+}^{\text{A}}(a_{i})_{b_{1}}  &  =p_{+}^{\text{A}}(a_{i})_{b_{2}%
},\label{eq NS1}\\
p_{+}^{\text{B}}(b_{j})_{a_{1}}  &  =p_{+}^{\text{B}}(b_{j})_{a_{2}},
\label{eq NS2}%
\end{align}
for $i,j\in\{1,2\}$. Ignoring the conditioning on the distant setting (due to
locality) and dropping the index $+$ everywhere, inequality (\ref{eq J new})
becomes the CH inequality \cite{Clau1974}:%
\begin{align}
C_{H}  &  \equiv+\ p(a_{1}b_{1})+p(a_{1}b_{2})+p(a_{2}b_{1})\nonumber\\
&  \ \ \ \ -p(a_{2}b_{2})-p^{\text{A}}(a_{1})-p^{\text{B}}(b_{1})\leq0.
\label{eq CH}%
\end{align}
Eberhard's main contribution was to realize that non-maximally entangled
states allow a violation of the CH or Eberhard inequality for detection
efficiencies as low as $2/3$, which is still the lowest known value for qubit
systems. In contrast, efficiency of $82.8\,\%$ is required for maximally
entangled states\ \cite{Garg1987,Lars1998}. The use of the CH or Eberhard
inequalities and non-maximally entangled states hence greatly eases the
detection efficiency requirements, one of the most challenging aspects of
photonic experiments. (We mention that there are also forms of the CH or
Eberhard inequality where all terms are divided by the sum of singles
probabilities or counts\ \cite{Chri2013,Khre2014}.)

The inequality (\ref{eq J new}), which we call the CH-E inequality, will be
used in the later sections, as it is the simplest known form, with only four
terms that all stem from mutually exclusive setting combinations.

\section{Space-time arrangement and setting predictability}

For a photonic Bell test, consider the space-time diagram in
Fig.\ \ref{Fig_Space-time_diagram}, where intervals of space-time events are
denoted with (non-italic) bold letters. A photon pair is emitted by a source
at $\mathbf{E}$. The photons travel a distance $d$ in fibers (solid blue
lines) with refractive index $n$ to Alice and Bob, where they pass the setting
devices, indicated by black rectangles. Geometric deviations from a perfectly
one-dimensional setup (black dashed lines) and any other additional delays are
represented by $\tau_{\text{G}}$. Alice's and Bob's measurement outcomes are
restricted to intervals $\mathbf{A}$ and $\mathbf{B}$ of duration
$\tau_{\text{M}}$. Outcome independence requires space-like separation of
$\mathbf{A}$ and $\mathbf{B}$. Setting independence requires that Alice's
setting generation is confined to interval $\mathbf{a}$, space-like separated
from Bob's outcome interval $\mathbf{B}$, and likewise $\mathbf{b}$ must be
space-like separated from $\mathbf{A}$. Space-like separation of the setting
generations within $\mathbf{a}$ and $\mathbf{b}$ from the emission interval
$\mathbf{E}$ closes the freedom-of-choice loophole. (The relevant space-like
separations can only be achieved by using at least three distinct locations.
One measurement device may be located at the source \cite{Sche2010,Erve2014},
but then the corresponding setting generator needs to be placed at a
distance.) The time duration $\tau_{\text{S}}$ for $\mathbf{a}$ and
$\mathbf{b}$ must be smaller than $\tau_{1}$, and the time for setting
generation as well as deployment of the setting (duration $\tau_{\text{D}}$)
must be smaller than $\tau_{2}$:\begin{figure}[tb]
\begin{center}
\includegraphics[width=0.48\textwidth]{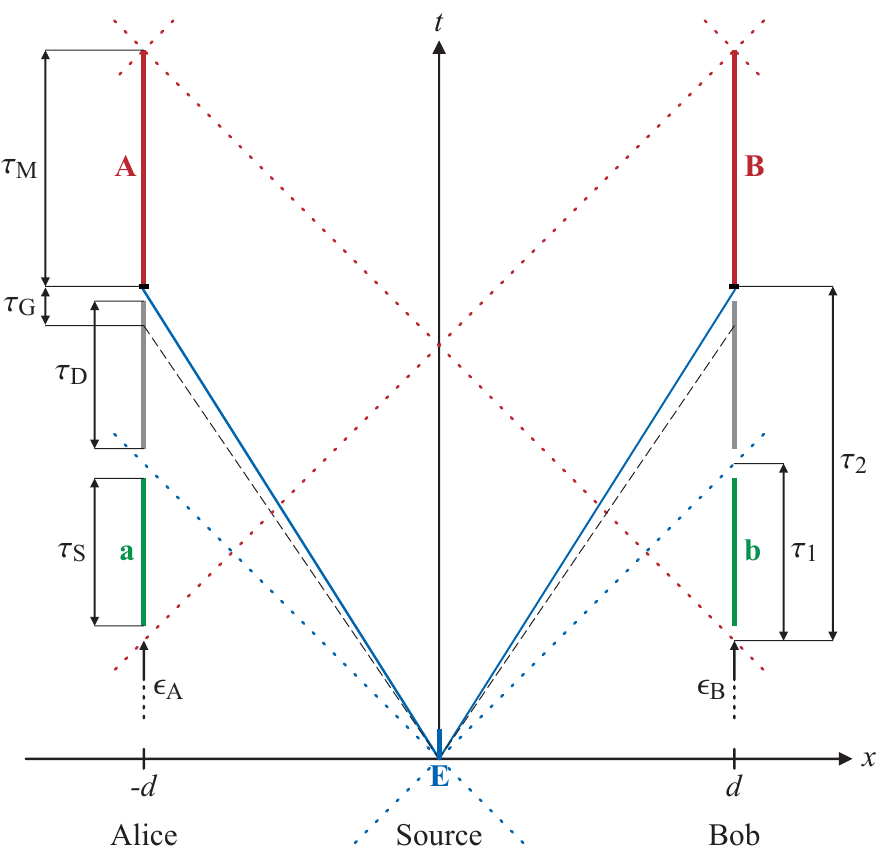}
\end{center}
\caption{Space-time diagram of a photonic fiber-based Bell test. \textbf{E}
represents the emission of a photon pair, \textbf{A} and \textbf{B} are
Alice's and Bob's detection intervals, and \textbf{a} and \textbf{b} are their
setting choice intervals. Relevant light cones are indicated by dotted lines.
Knowledge about the distant setting that can be available at Bob's (Alice's)
measurement device is quantified by $\epsilon_{\text{A}}$ ($\epsilon
_{\text{B}}$). See main text for further details.}%
\label{Fig_Space-time_diagram}%
\end{figure}%
\begin{align}
\tau_{\text{S}}  &  <\tau_{1}=\frac{(3-n)\,d}{c_{0}}-(\tau_{\text{G}}%
+\tau_{\text{M}}),\\
\tau_{\text{S}}+\tau_{\text{D}}  &  <\tau_{2}=\frac{2\,d}{c_{0}}%
-\tau_{\text{M}}.
\end{align}
Here, $c_{0}$ denotes the speed of light in vacuum.

Closing the locality and freedom-of-choice loopholes requires generation of
fast random numbers for the settings $(a,b)$ which must not be able to
influence the respective distant outcome (locality) or have a mutual
interdependence with the hidden variable $\lambda$ (freedom of choice). It
should be noted, however, that the requirements for $(a,b)$-$\lambda$
independence in a Bell test differ in important ways from \textquotedblleft
randomness\textquotedblright\ as per the usual definitions. For example, it is
common to consider as random a source of independent, identically-distributed,
unbiased bits $x_{i}$, described by the probabilities $p(x_{i}|x_{j\neq
i})=\frac{1}{2}$. Using such sources to choose $(a,b)$ does not by itself
guarantee independence from $\lambda$, because $\lambda$ could influence $x$
in such a way that $x$ is predictable knowing $\lambda$, but fully
unpredictable absent this knowledge. In contrast, a source that is biased but
uninfluenced by $\lambda$, e.g.\ $p(x_{i}|\lambda)=\frac{3}{4}$, is suitable
for generating the required independence, despite being far from random by the
usual definitions.

As concerns physical variables and setting choices, we use the term
\textquotedblleft random\textquotedblright\ to mean independence from
$\lambda$. Physically, this independence can be compromised by an influence of
$\lambda$ on $x$, by an influence of $x$ on $\lambda$, or by a common
influence. The first two of these can be excluded by space-like separation of
the setting generation from the creation of the hidden variables, while the
last one is excluded if $\lambda$ and/or $x$ is uninfluenced, i.e.\ stochastic.

An entire Bell experiment, including the setting generation, must be viewed
within local realism, and quantum mechanics must not be invoked. Candidate
stochastic processes include chaotic dynamics, human
decision-making\ \cite{Bell2004}, and cosmic light sources\ \cite{Gall2014}.
Photonic devices use the reflection/transmission at a beam
splitter\ \cite{Jenn2000} or the emission/detection time \cite{Fuer2010},
population \cite{Stip2015}, or phase \cite{Abel2014,Abel2015} of a coherent
light source. It bears repeating that a local realist model must contain some
stochastic element if it is to be testable.

Any real implementation of a random number generator will to some extent be
influenced by effects prior to the generation, giving non-zero predictive
power beyond the \textit{a priori} probability of guessing the eventual
setting. We call this the \textit{excess predictability}. This opens the
locality loophole or freedom-of-choice loophole to some extent. In general,
each setting choice could have a different excess predictability in every
trial, such that in trial $n$ the excess predictability takes on values
$\epsilon_{\text{A}}^{(i)}$ and $\epsilon_{\text{B}}^{(i)}$ with
$\epsilon_{\text{A}}^{(i)},\epsilon_{\text{B}}^{(i)}\in\lbrack0,1]$. Below we
model two special cases: We either assume that in a small \textit{fraction}
$\epsilon_{\text{A}}$ ($\epsilon_{\text{B}}$) of experimental runs Alice's
(Bob's) setting choice is \textit{perfectly} communicable to the distant party
Bob (Alice), or we consider that in \textit{every} trial Bob (Alice) can
predict the distant setting with a small certainty $\epsilon_{\text{A}}$
($\epsilon_{\text{B}}$) better than the \textit{a priori} probability.
Presumably, the physical situation could be any mixture of these two models.

\section{Adaptation of the CH-E inequality}

To use the CH-E inequality, which employs conditional probabilities, we need
the concept of a trial. Without an exact definition of what a trial is, it is
unclear how to use normalized counts or the concept of probabilities when
employing the CH-E\ inequality. Normalization with respect to the pair
production rate or measurement time for a given fixed setting
\cite{Kofl2013,Khre2014} will not be possible for a loophole-free Bell test
because the analysis technique for closing the memory loophole relies on the
concept of trials. Noting that the particular construction and assumptions
involved in a given test might refine the operational definition of a trial in
that test, we suggest that the reader consider a trial most basically as a
(locally-defined) measurement interval, for which each measurement party must
record exactly one outcome (possibly including \textquotedblleft undetected").

Specifically, we have in mind a pulsed experiment, where every pulse -- which
might or might not create a down-conversion pair -- belongs to exactly one
trial. We will not consider anything that happens between the trials. Fixed
measurement time windows synchronized with the laser pulses are also suitable
for closing the coincidence-time loophole for the CH inequality
\cite{Lars2013}.

Given that information about a setting will sometimes exist in the backward
light cone of the distant outcome event, it is necessary to adapt the CH-E
inequality. We now consider two different mathematical models for the
communication or excess predictability of the setting values:

\textit{Scenario (i) -- communication in some trials}. Here, in a fraction
$\epsilon_{\text{A}}$ ($\epsilon_{\text{B}}$) of the trials, Alice's (Bob's)
setting is perfectly known to Bob (Alice) via communication, while in the rest
of the trials the locality condition is perfectly fulfilled. For simplicity,
we assume that this fraction is the same for all setting combinations. To be
conservative, we shall not assume that the \textquotedblleft glitches" of too
early settings happen statistically independently on the two sides, but that
they may avoid happening in the same trials. We introduce the abbreviation
\begin{equation}
\epsilon_{\text{AB}}\equiv\min(\epsilon_{\text{A}}\!+\!\epsilon_{\text{B}},1)
\label{eq epsAB}%
\end{equation}
for the (maximal possible) fraction where one setting is communicable to the
distant outcome. Let us consider the subset $S_{\!\text{A}}$ of trials in
which Alice's setting $a$ is communicated to Bob's measurement device while
her measurement device has no information about Bob's setting $b$. It is
conceivable that Alice's devices know when her setting is communicated. Then
the strategy is as follows: Alice's measurement device \textquotedblleft
overrules" whatever fate has been designated and outputs $+$. Bob also outputs
$+$, unless $a=a_{2}$ and $b=b_{2}$, whereupon he outputs $0$. For the
different setting combinations, their measurement results therefore contribute
to $p_{++}(a_{1}b_{1})$, $p_{++}(a_{1}b_{2})$, $p_{++}(a_{2}b_{1})$, and
$p_{+0}(a_{2}b_{2})$, and nothing else. The last three terms do not appear in
the CH-E inequality (\ref{eq J new}), and the first is beneficial for its
violation. The $J$ value in the subset $S_{\!\text{A}}$ can therefore reach
the logical bound $J=+1$. Importantly, also those events that would have had
fate $00$ contributed to the violation.

Straightforwardly, the above arguments can be repeated for the subset
$S_{\!\text{B}}$ of trials where Bob's setting can be communicated but not
Alice's and for the subset $S_{\!\text{AB}}$ where both can be communicated.
This implies that local hidden variables augmented with setting communication
can attain the CH-E value $+1$ in the total subset $S_{\!\epsilon_{\text{AB}}%
}=S_{\!\text{A}}\cup S_{\!\text{B}}\cup S_{\!\text{AB}}$ whose size is bounded
by the fraction $\epsilon_{\text{AB}}$ of all trials. This means that for the
entirety of all trials such models reach $J=\epsilon_{\text{AB}}$. The CH-E
inequality $J\leq0$ must therefore be rewritten with an \textit{adapted
bound}:%
\begin{equation}
J\leq\epsilon_{\text{AB}}. \label{eq J adapt}%
\end{equation}
In other words, when physical (sub)luminal communication of a setting to a
distant outcome is possible in a fraction $\epsilon_{\text{AB}}$ of trials,
the collected results must violate inequality\ (\ref{eq J adapt}) with its
adapted bound to rule out a local realist explanation.

An important remark: The above strategy violates the no-signaling condition
(\ref{eq NS2}). From subset $S_{\!\text{A}}$ one has contributions to the
singles probability $p_{+}^{\text{B}}(b_{2})_{a_{1}}$ but not to
$p_{+}^{\text{B}}(b_{2})_{a_{2}}$. This violation is a general feature of pure
strategies with communication. Mixed strategies can hide the communication and
obey no-signaling. When the entire setting information is communicated, the
predictions of every no-one-way-signaling distribution can be simulated by
local hidden variables \cite{Baco2003}. The optimal no-signaling strategy is
the simulation of a PR box~\cite{Pope1994}, which works as follows: For every
trial, Alice and Bob share a random variable $r\in\{+,0\}$ with distribution
$p(r\!=\!+)=p(r\!=\!0)=\frac{1}{2}$. When Alice transmits her setting $a$ to
Bob, she outputs $A=r$. Bob also outputs $B=r$ unless $a=a_{2}$ and $b=b_{2}$
in which case he produces the opposite result ($+$ if $r=0$, $0$ if $r=+$).
This strategy obeys no-signaling and, within the subset $S_{\!\text{A}}$,
reaches $J=\frac{1}{2}$. Note that for the CHSH inequality the logical and the
no-signaling bound are identical (equal to 4). This is not the case for the
CH-E inequality, where the logical bound is $1$ and the no-signaling bound is
$\frac{1}{2}$.

While the bound $\epsilon_{\text{AB}}$ in (\ref{eq J adapt}) cannot be reached
by local hidden variable models that are augmented by setting communication
and obey the no-signaling conditions, the bound is conservative only by a
factor of 2 (since according to the above, the bound for communication
strategies obeying no-signaling is $\frac{\epsilon_{\text{AB}}}{2}$).
Moreover, it has the advantage that one need not additionally check the
no-signaling conditions in an experiment. Having quantified $\epsilon
_{\text{AB}}$, one can solely rely on the inequality (\ref{eq J adapt})
itself. Also note that violation of the no-signaling conditions within the
subensemble $S_{\!\epsilon_{\text{AB}}}$ could be due to actual (sub)luminal
signals and would not be in contradiction with causality.

\textit{Scenario (ii) -- excess predictability in all trials}. In this
scenario, we assume that, in every run, Alice's and Bob's setting choices $a$
and $b$ are partially dependent on external influences that are available also
at the distant measurement event. Formally, this corresponds to a violation of
the locality and freedom-of-choice assumptions. We can incorporate all these
influences together with the properties $\lambda$ of the photon pair into a
joint set $\mu$ of hidden variables. However, similar to Ref.\ \cite{Puet2014}
we assume that in every run the effect of $\mu$ cannot alter the probability
for a specific setting choice by more than a certain number, quantified by
parameters $\epsilon_{\text{A}}$ and $\epsilon_{\text{B}}$ in the following
way:%
\begin{align}
(1-\epsilon_{\text{A}})\,p(a)  &  \leq p(a|\mu)\leq(1+\epsilon_{\text{A}%
})\,p(a)\label{eq P(a)}\\
(1-\epsilon_{\text{B}})\,p(b)  &  \leq p(b|\mu)\leq(1+\epsilon_{\text{B}%
})\,p(b) \label{eq P(b)}%
\end{align}
Using $p(a,b|\mu)=p(a|\mu)\,p(b|\mu)$, which is guaranteed as $\mu$ carries
all hidden properties, and abbreviating%
\begin{equation}
\epsilon_{\pm}\equiv\epsilon_{\text{A}}+\epsilon_{\text{B}}\pm\epsilon
_{\text{A}}\epsilon_{\text{B}}, \label{eq epspm}%
\end{equation}
we obtain%
\begin{equation}
(1-\epsilon_{_{-}})\,p(a)\,p(b)\leq p(a,b|\mu)\leq(1+\epsilon_{+}%
)\,p(a)\,p(b). \label{eq P(a,b|mu)}%
\end{equation}
Zero excess predictability implies $p(a,b)=p(a)\,p(b)$, while the converse is
not true. Note that the individual setting probabilities $p(a)$ and $p(b)$ can
have non-zero biases $\kappa_{\text{A}},\kappa_{\text{B}}\in(-\frac{1}%
{2},\frac{1}{2})$,%
\begin{align}
p(a_{1})  &  =\tfrac{1}{2}-\kappa_{\text{A}},\;p(a_{2})=\tfrac{1}{2}%
+\kappa_{\text{A}},\label{eq bias A}\\
p(b_{1})  &  =\tfrac{1}{2}-\kappa_{\text{B}},\;p(b_{2})=\tfrac{1}{2}%
+\kappa_{\text{B}}. \label{eq bias B}%
\end{align}
which are neither at variance with the locality or freedom-of-choice
assumptions nor problematic in the derivation of the CH-E inequality. The
parameters $\epsilon_{\text{A}}$ and $\epsilon_{\text{B}}$ in (\ref{eq P(a)})
and (\ref{eq P(b)}) hence quantify \textit{predictability beyond bias}.

Recorded data allows us to estimate total probabilities averaged over $\mu$,
that is, $p(A,B,a,b)=\;$E$[p(A,B,a,b|\mu)]$, with E denoting the expectation
value, but does not immediately allow us to estimate the conditional
probabilities $p_{AB}(ab)\equiv p(A,B|a,b)$. The latter are well-defined for
each individual value of $\mu$, which is inaccessible to us. When conditioned
on $\mu$, the conditional probabilities obey the CH-E inequality
(\ref{eq J new}):%
\begin{align}
&  +p(\text{++}|a_{1}b_{1},\mu)-p(\text{+0}|a_{1}b_{2},\mu)\nonumber\\
&  -p(\text{0+}|a_{2}b_{1},\mu)-p(\text{++}|a_{2}b_{2},\mu)\leq0.
\label{eq CH-E mu}%
\end{align}
Using (\ref{eq P(a,b|mu)}), we obtain%
\begin{equation}
\frac{p(A,B,a,b|\mu)}{p(a)\,p(b)\,(1+\epsilon_{+})}\leq\frac{p(A,B,a,b|\mu
)}{p(a,b|\mu)}\leq\frac{p(A,B,a,b|\mu)}{p(a)\,p(b)\,(1-\epsilon_{_{-}})},
\label{eq P-ineq}%
\end{equation}
The inequalities (\ref{eq P-ineq}) must also hold for expectation values:%
\begin{equation}
\frac{p(A,B,a,b)}{p(a)\,p(b)\,(1+\epsilon_{+})}\leq\text{E}[p(A,B|a,b,\mu
)]\leq\frac{p(A,B,a,b)}{p(a)\,p(b)\,(1-\epsilon_{_{-}})},
\label{eq bounded E[p]}%
\end{equation}
where $p(A,B|a,b,\mu)=\frac{p(A,B,a,b|\mu)}{p(a,b|\mu)}$. This allows us to
arrive at the following \textit{adapted form} of the CH-E inequality:%
\begin{align}
J_{\epsilon}\equiv &  +\frac{p(\text{++},a_{1}b_{1})}{p(a_{1})\,p(b_{1}%
)\,(1+\epsilon_{+})}-\frac{p(\text{+0},a_{1}b_{2})}{p(a_{1})\,p(b_{2}%
)\,(1-\epsilon_{_{-}})}\nonumber\\
&  -\frac{p(\text{0+},a_{2}b_{1})}{p(a_{2})\,p(b_{1})\,(1-\epsilon_{_{-}}%
)}-\frac{p(\text{++},a_{2}b_{2})}{p(a_{2})\,p(b_{2})\,(1-\epsilon_{_{-}})}%
\leq0. \label{eq adapted CH-E}%
\end{align}
The inequality holds because, due to (\ref{eq bounded E[p]}), the left hand
side is bounded by E$[p($++$|a_{1}b_{1},\mu)-p($+0$|a_{1}b_{2},\mu
)-p($0+$|a_{2}b_{1},\mu)-p($++$|a_{2}b_{2},\mu)]$, which, due to
(\ref{eq CH-E mu}), is bounded by 0.

It is important to note that the adaptation in scenario (i) is
\textquotedblleft absolute\textquotedblright, while the one in scenario (ii)
is \textquotedblleft fractional\textquotedblright. The adapted bound violation
from a given measured $J$ can withstand a larger value of $\epsilon_{\pm}$ in
scenario (ii) than $\epsilon_{\text{AB}}$ in scenario (i).

We conclude that the concrete adapted form of the CH-E inequality depends on
the physical scenario of how setting choices are communicable to or
predictable at the remote side.

\section{Statistical significance and run time}

In published experimental tests of Bell's inequality, it is common to report a
violation as the number of standard deviations separating the measured value
from the local realist bound, assuming Poissonian statistics. This quantifies
the chance that a value consistent with local realism is still in agreement
with the experimental data. In fact, we are interested in a different
question: What is the chance that the violation observed in the experiment
could have been produced under local realism? Moreover, to close the memory
loophole, we may no longer assume that the trials are i.i.d. Employing the
concept of hypothesis testing, for instance using the Hoeffding inequality
\cite{Hoef1963}, one can put a bound on the probability that local realism
produced the data in a given experiment, even when allowing memory.

Based on the works\ \cite{Gill2003b,Bier2015}, we present the first
statistical analysis with the following three key features, all of which are
essential for a photonic Bell state with current technology:

\begin{enumerate}
\item We allow for a bias in the setting choices.

\item We take into account a communication or excess predictability (beyond
bias) of the setting choices, using adapted versions of the CH-E inequality.

\item We apply Doob's theorem to get rid of non-contributing trials and reduce
the experimental run time to an acceptable level.
\end{enumerate}

While all points are well understood individually, point 3 becomes non-trivial
when combined with 1 and 2.

A \textit{supermartingale} is a stochastic process for which, at any time in
the sequence, the expectation value of the next value in the sequence does not
exceed the expectation value of the current value in the sequence, given
knowledge of all of measurements in the history of the process. (One can think
of it as a random walk with memory and strictly non-positive drift.)

We consider an experiment with $N$ trials. In each trial $n=1,\ldots,N$, a
measurement involves choosing a pair of settings and recording an outcome for
each party, leading to an experimental value $\Delta^{(n)}$ for that trial,
according to the inequality. Consider the random process $Z_{\Delta}$:
$Z^{(0)}\!=\!0,Z^{(1)},...,Z^{(N)}$ with $Z^{(l)}\!=\!%
{\textstyle\sum\nolimits_{n=1}^{l}}
\Delta^{(n)}$, whose increments $\Delta^{(n)}$ fall within range $r_{\Delta}$.
Then, Hoeffding's inequality%
\begin{equation}
p\Big(Z^{(N)}-\text{E}[Z^{(N)}]\geq c\sqrt{N}\Big)\leq\text{e}^{-\frac
{2}{r_{\Delta}^{2}}c^{2}} \label{eq Hoeffding Z}%
\end{equation}
bounds the probability that, after $N$ trials, $Z^{(N)}$ can exceed the value
E$[Z^{(N)}]+c\sqrt{N}$, where $c$ is a positive number and E$[Z^{(N)}]$ is the
expectation value of $Z^{(N)}$. Inequality (\ref{eq Hoeffding Z}) holds for
i.i.d.\ trials but also (in the weaker case) when $Z^{(N)}$ is a
supermartingale, i.e.\ E$[Z^{(l+1)}|Z^{(1)},...,Z^{(l)}]\leq Z^{(l)}$ for all
$l$, or equivalently E$[\Delta^{(n)}|\Delta^{(1)},...,\Delta^{(n-1)}]\leq0$
for all $n$.

We will examine separately the case where local realism (LR) holds fully, the
case where local realism fails in the way described in scenario (i) in section
V, and the case where it fails in the way described in scenario (ii) in
section V. We call the latter two situations \textquotedblleft$\epsilon$ local
realism" ($\epsilon$LR).%
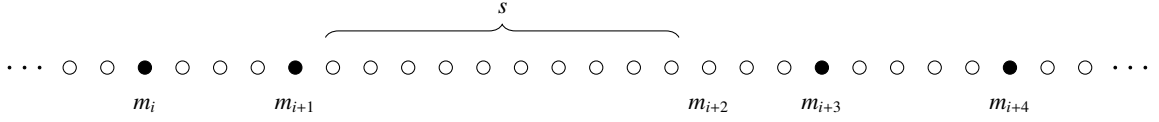
\begin{figure*}[t]%
\begin{tikzpicture}
\draw [fill=black] (-8.3,0) circle (0.5pt); \draw [fill=black]
(-8.1,0) circle (0.5pt); \draw [fill=black] (-7.9,0) circle (0.5pt);
\draw (-7.5,0) circle (2.5pt); \draw (-7.0,0) circle (2.5pt); \draw
[fill=black] (-6.5,0) circle (2.5pt); \coordinate[label=center:$m_i$]
(i) at (-6.5,-0.5); \draw (-6.0,0) circle (2.5pt); \draw (-5.5,0)
circle (2.5pt); \draw (-5.0,0) circle (2.5pt); \draw [fill=black]
(-4.5,0) circle (2.5pt); \coordinate[label=center:$m_{i+1}$] (i+1)
at (-4.5,-0.5); \draw (-4.0,0) circle (2.5pt); \draw (-3.5,0) circle
(2.5pt); \draw (-3.0,0) circle (2.5pt); \draw (-2.5,0) circle
(2.5pt); \draw (-2.0,0) circle (2.5pt); \draw (-1.5,0) circle
(2.5pt); \draw (-1.0,0) circle (2.5pt); \draw (-0.5,0) circle
(2.5pt); \draw (0.0,0) circle (2.5pt); \draw (0.5,0) circle (2.5pt);
\draw (1.0,0) circle (2.5pt); \coordinate[label=center:$m_{i+2}$]
(i+2) at (1.0,-0.5); \draw (1.5,0) circle (2.5pt); \draw (2.0,0)
circle (2.5pt); \draw [fill=black] (2.5,0) circle (2.5pt);
\coordinate[label=center:$m_{i+3}$] (i+3) at (2.5,-0.5); \draw
(3.0,0) circle (2.5pt); \draw (3.5,0) circle (2.5pt); \draw (4.0,0)
circle (2.5pt); \draw (4.5,0) circle (2.5pt); \draw [fill=black]
(5.0,0) circle (2.5pt); \coordinate[label=center:$m_{i+4}$] (i+4) at
(5.0,-0.5); \draw (5.5,0) circle (2.5pt); \draw (6.0,0) circle
(2.5pt); \draw [fill=black] (6.4,0) circle (0.5pt); \draw
[fill=black] (6.6,0) circle (0.5pt); \draw [fill=black] (6.8,0)
circle (0.5pt); \draw [decorate,decoration={brace,amplitude=05pt}]
(-4.1,0.4) -- (0.6,0.4); \node at (-1.75,0.8) {$s$};
\end{tikzpicture}\caption{Illustration of the dilution scheme. The circles represent
the experimental trials $n=1,...,N$, where white and black fillings
correspond to the values $K^{(n)}=-\epsilon_{\text{AB}}$ and
$K^{(n)}\neq-\epsilon_{\text{AB}}$, respectively. The concentrated process has ``stopping times" $m_i$, which encompass all black trials as well as all those, including white, which are preceded by a streak of $s$ subsequent occurrences of white trials.\label{fig_doob}}%
\end{figure*}%

\textit{Under local realism:} Consider the random process $Z_{J}$:
$Z_{J}^{(0)}\!=\!0,Z_{J}^{(1)},...,Z_{J}^{(N)}$ with $Z_{J}^{(l)}\!=\!%
{\textstyle\sum\nolimits_{n=1}^{l}}
J^{(n)}$, where the measured value (i.e.\ the increment of the process) in run
$n$ is denoted by $J^{(n)}$. We abbreviate $p_{ij}\equiv p(a_{i})\,p(b_{j})$
which, under freedom of choice, equals $p(a_{i}b_{j})$, i.e.\ the probability
that Alice chooses setting $a_{i}$ and Bob chooses $b_{j}$. Due to the setting
biases these 4 values need not be $\frac{1}{4}$. Furthermore, we label with
$X_{ij}^{AB}$ those trials where Alice chooses setting $a_{i}$ and observes
outcome $A\in\{+,0\}$, and Bob chooses $b_{j}$ and observes outcome
$B\in\{+,0\}$. The increments $J^{(n)}$ are defined as%
\begin{equation}
J^{(n)}\equiv\left\{
\begin{array}
[c]{rr}%
+\frac{1}{p_{11}} & \;\text{for }X_{11}^{++}\\
-\frac{1}{p_{12}} & \text{for }X_{12}^{+0}\\
-\frac{1}{p_{21}} & \text{for }X_{21}^{0+}\\
-\frac{1}{p_{22}} & \text{for }X_{22}^{++}\\
0 & \text{else}%
\end{array}
\right.  \label{eq J(n)}%
\end{equation}
The probability for a trial $X_{ij}^{AB}$ is given by the probability $p_{ij}$
that the setting combination $a_{i}b_{j}$ is chosen, multiplied with the
conditional probability to observe the outcomes $A$ and $B$ given this setting
choice: $p(X_{ij}^{AB})=p_{ij}\,p_{AB}(a_{i},b_{j})$. The definition
(\ref{eq J(n)}) thus assures that the expectation value of $J^{(n)}$ is
precisely given by $J$ from (\ref{eq J new}) and hence, under local realism,
is bounded by zero. (Note that, unlike $J$, the process $Z_{J}$ scales with
$N$ unboundedly.) Even allowing memory, the expected value of every increment
is still bounded by zero: E$[J^{(n)}|J^{(1)},...,J^{(n-1)}]\leq0$, making the
process $Z_{J}$ a supermartingale. The increments fall within the range%
\begin{equation}
r_{J}=\tfrac{1}{p_{11}}+\max(\tfrac{1}{p_{12}},\tfrac{1}{p_{21}},\tfrac
{1}{p_{22}}),
\end{equation}
which is close to 8 for small biases. The Hoeffding inequality for $Z_{J}$
reads%
\begin{equation}
p_{\text{LR}}\Big(Z_{J}^{(N)}\geq c\sqrt{N}\Big)\leq\text{e}^{-\frac{2}%
{r_{J}^{2}}c^{2}}. \label{eq Hoeffding J}%
\end{equation}

\textit{Scenario (i) -- communication in some trials}. Now we consider the
case of $\epsilon$LR in the scenario (i) of section V, i.e.\ the adapted
inequality (\ref{eq J adapt}). If LR fails altogether, the expectation value
of $J^{(n)}$ can reach $1$. If LR fails only due rare communication events,
and if we assume these failures are independent of the history of the
experiment, then the expectation value of $J^{(n)}$ can reach $\epsilon
_{\text{AB}}$. This means that under $\epsilon$LR, $Z_{J}$ is no longer a
supermartingale. We define the process $Z_{K}$ with increments%
\begin{equation}
K^{(n)}\equiv J^{(n)}-\epsilon_{\text{AB}}, \label{eq K J}%
\end{equation}
Due to E$[K^{(n)}|K^{(1)},...,K^{(n-1)}]\leq0$ the process $Z_{K}$ is a
supermartingale also in scenario-(i) $\epsilon$LR. The trial values $K^{(n)}$
still have range $r_{J}$. The Hoeffding inequality then reads%
\begin{equation}
p_{\epsilon\text{LR}}\Big(Z_{K}^{(N)}\geq c\sqrt{N}\Big)\leq\text{e}%
^{-\frac{2}{r_{J}^{2}}c^{2}}, \label{eq Hoeffding K}%
\end{equation}
where, using eq.\ (\ref{eq K J}), one can replace $Z_{K}^{(N)}$ by
$Z_{J}^{(N)}-N\,\epsilon_{\text{AB}}$.

If we denote by $R$ the frequency of trials and by $J$\ the experimentally
expected value, then, assuming small bias, the condition $Z_{J}^{(N)}\geq
N\,\epsilon_{\text{AB}}+c\sqrt{N}$ in (\ref{eq Hoeffding K}) is likely to be
reached after a run time of $\frac{c^{2}}{R\,(J-\epsilon_{\text{AB}})^{2}}$.
In a photonic Bell experiment with total collection efficiency $\eta
\approx75\,\%$\ \cite{Gius2013,Chri2013}, one down-conversion pair in $10^{3}$
pulses, and reasonable state visibility and rate of dark/background counts,
the CH-E value would be of the order of $J\sim10^{-6}$. (The low probability
for a pair-production dominates, but the state and measurement angles used at
this detection efficiency also contribute to the smallness of this number.)
Assuming a pulse rate of $R\approx1\,$MHz, $\epsilon_{\text{AB}}\approx
10^{-7}$, and $r_{J}\approx8$, particle-physics \textquotedblleft gold
standard" significance of $p\sim10^{-6}$ (i.e.\ $c\approx20$) would only be
reached after a run time of approximately 16 years, which exceeds the average
duration of PhD studies.

Fortunately, however, this result can be improved using Doob's optional
stopping theorem. Following Ref.\ \cite{Gill2003b}, we first estimate the
fraction of all trials $n$ for which the $J^{(n)}$ value is non-zero:%
\begin{equation}
f=\frac{\sharp\{n\!:J^{(n)}\neq0\}}{N}.
\end{equation}
By inspection of (\ref{eq J}), these are the trials $X_{11}^{++},X_{12}%
^{+0},X_{21}^{0+},X_{22}^{++}$. All other combinations of settings and
outcomes do not contribute to the CH-E value, i.e.\ have $J^{(n)}=0$ and hence
$K^{(n)}=-\epsilon_{\text{AB}}$. With the experimental parameters from above,
we estimate that the fraction of contributing trials is $f\approx2\cdot
10^{-5}$.

With Doob's optional stopping theorem it is possible to increase the
statistical significance of a given data set by looking at a \textquotedblleft
concentrated process". If $Z_{J}$ were a supermartingale, which it is only in
LR and not in scenario-(i) $\epsilon$LR, then the procedure would be rather
straightforward as one could simply skip all non-contributing trials with
$J^{(n)}=0$. Our case is more complicated, as those non-contributing trials
have (negative) value\ $K^{(n)}=-\epsilon_{\text{AB}}$ and hence do in fact
contribute to $Z_{K}$.

We propose the following solution to this problem (see Fig.\ \ref{fig_doob}):
Let us consider the aggregated value $Z_{K}^{(m)}=%
{\textstyle\sum\nolimits_{n=1}^{m}}
K^{(n)}$ at $M$ specific \textquotedblleft stopping times" $m\in\{m_{1}%
,m_{2},...,m_{M}\}$, namely those where (a) $K^{(m)}\neq-\epsilon_{\text{AB}}$
or (b) $K^{(m)}=-\epsilon_{\text{AB}}$ when preceded by a \textquotedblleft
streak" of $s$ contiguous occurrences of $K^{(n=m-s,...,m-1)}=-\epsilon
_{\text{AB}}$. Every stop starts a fresh streak. Let us abbreviate $\frac
{1}{w}=\max(\tfrac{1}{p_{12}},\tfrac{1}{p_{21}},\tfrac{1}{p_{22}})$. This
choice of $m$ ensures, without looking into the future, that the increment
from any $Z_{K}^{(m_{i})}$ to $Z_{K}^{(m_{i+1})}$ is between $-\frac{1}%
{w}-(s\!+\!1)\epsilon_{\text{AB}}$ (which one gets for $m_{i+1}=m_{i}+s+1$
when there is a streak of $s$ occurrences of $K^{(n=m_{i}+1,...,m_{i+1}%
-1)}=-\epsilon_{\text{AB}}$ and then a final $K^{(n=m_{i+1})}=-\frac{1}%
{w}-\epsilon_{\text{AB}}$) and $\frac{1}{p_{11}}-\epsilon_{\text{AB}}$ (which
one gets for $m_{i+1}=m_{i}+1$, and $K^{(n=m_{i+1})}=\frac{1}{p_{11}}%
-\epsilon_{\text{AB}}$). This implies that the concentrated process
$Z_{K}^{(m_{1})},Z_{K}^{(m_{2})},...,Z_{K}^{(m_{M})}$ is a supermartingale
with range%
\begin{equation}
r_{J,s}=r_{J}+s\,\epsilon_{\text{AB}}.
\end{equation}
The Hoeffding inequality (\ref{eq Hoeffding K}) is now altered in two ways.
First, the range increases from $r_{J}$ to $r_{J,s}$. Second, in the
concentrated process \textquotedblleft time is now running faster"
\cite{Gill2003b} which means that $N$ gets replaced by the concentrated
process length $M$, which is the number of stopping times above. Hence,%
\begin{equation}
p_{\epsilon\text{LR},M}\Big(Z_{K}^{(m_{M})}\geq c\sqrt{M}\Big)\leq
\text{e}^{-\frac{2}{r_{J,s}^{2}}c^{2}}. \label{eq Hoeffding ZM}%
\end{equation}
Note that with $s=0$ one recovers the original process, i.e.\ $r_{J,s=0}%
=r_{J}$ and $M=N$, and thus ineq.\ (\ref{eq Hoeffding K}). Using
eq.\ (\ref{eq K J}), $Z_{K}^{(m_{M})}$ can be replaced by $Z_{J}^{(m_{M}%
)}-m_{M}\,\epsilon_{\text{AB}}$. When most of the trials are non-contributing,
one can choose $s$ such that $M\ll N$ while $m_{M}\approx N$.

Now we focus our attention again to an estimation of the experimental run
time. For our purposes it is not necessary to find the optimal value for $s$,
which will in general depend on $f$ and $\epsilon_{\text{AB}}$. We will see
the remarkable power of Doob's theorem already by choosing $s=\lfloor
\epsilon_{\text{AB}}^{-1}\rfloor$ which, for small biases and small
$\epsilon_{\text{AB}}$, leads to range $r_{J,s}\approx9$. Because of
$f\gg\epsilon_{\text{AB}}$ it (almost) never happens that there are full
streaks of $\lfloor\epsilon_{\text{AB}}^{-1}\rfloor\!+\!1$ subsequent
occurrences of non-contributing $J^{(n)}$ trials. Hence we can take $M\approx
fN$, meaning that the concentrated process stops at (almost) exactly the
contributing trials. The condition in (\ref{eq Hoeffding ZM}) is likely to be
reached after a run time of $\frac{c^{2}\,f}{R\,(J-\epsilon_{\text{AB}})^{2}}%
$. To obtain the same statistical significance as before ($p\sim10^{-6}$), we
now need to increase $c\approx20$ by a factor of $\frac{r_{J,s}}{r_{J}}%
\approx\frac{9}{8}$ to $c\approx22.5$. In total, Doob's theorem leads to a
remarkable reduction of the run time by a factor of $\frac{9^{2}}{8^{2}}f$
from 16 years to 3 hours, right into the range of experimental feasibility.
(We note that although the specific experimental values in a future optical
Bell test may differ substantially from our estimates, it is very likely that
within the near future the application of Doob's theorem as just outlined is
essential to achieve good statistical significance within a feasible run time.)

\textit{Scenario (ii) -- excess predictability in all trials}. We now consider
the case of $\epsilon$LR in scenario (ii) of section V, i.e.\ the adapted
inequality (\ref{eq adapted CH-E}). This situation is simpler than scenario
(i). We can define the increments of a process $Z_{J_{\epsilon}}$\ as%
\begin{equation}
J_{\epsilon}^{(n)}\equiv\left\{
\begin{array}
[c]{rr}%
+\frac{1}{p_{11}(1+\epsilon_{+})} & \;\text{for }X_{11}^{++}\\
-\frac{1}{p_{12}(1-\epsilon_{_{-}})} & \text{for }X_{12}^{+0}\\
-\frac{1}{p_{21}(1-\epsilon_{_{-}})} & \text{for }X_{21}^{0+}\\
-\frac{1}{p_{22}(1-\epsilon_{_{-}})} & \text{for }X_{22}^{++}\\
0 & \text{else}%
\end{array}
\right.  \label{eq Je(n)}%
\end{equation}
with $p_{ij}\equiv p(a_{i})\,p(b_{j})$ which need not equal $p(a_{i}b_{j})$.
This process has range%
\begin{equation}
r_{J_{\epsilon}}=\tfrac{1}{p_{11}(1+\epsilon_{_{+}})}+\max(\tfrac{1}{p_{12}%
},\tfrac{1}{p_{21}},\tfrac{1}{p_{22}})\tfrac{1}{1-\epsilon_{_{-}}},
\label{eq rke}%
\end{equation}
which, for small biases and small $\epsilon_{_{\pm}}$, is close to 8. Even
allowing memory, the expectation value of $J^{(n)}$ is precisely given by
$J_{\epsilon}$ from (\ref{eq adapted CH-E}), making the process
$Z_{J_{\epsilon}}$ a supermartingale. Doob's theorem can be applied right away
and all non-contributing trials can be discarded. With $M$ contributing
trials, the Hoeffding inequality reads%
\begin{equation}
p_{\epsilon\text{LR},M}\Big(Z_{J_{\epsilon}}^{(N)}\geq c\sqrt{M}%
\Big)\leq\text{e}^{-\frac{2}{r_{J_{\epsilon}}^{2}}c^{2}}. \label{eq peLR}%
\end{equation}

If the bound (\ref{eq P(a,b|mu)}) fails sometimes, say with probability
$q_{\text{f}}$, then the algebraic bound of $J_{\epsilon}$, which is
$\tfrac{1}{1-\epsilon_{_{-}}}$, can be reached in these trials. The above
formulas have to be adapted in the following way, using the logic from
scenario (i): A process $Z_{K_{\epsilon}}$ is defined such that
$Z_{K_{\epsilon}}^{(N)}=Z_{J_{\epsilon}}^{(N)}-N\,\tfrac{q_{\text{f}}%
}{1-\epsilon_{-}}$ with range $r_{K_{\epsilon}}+s\,\tfrac{q_{\text{f}}%
}{1-\epsilon_{-}}$, using streak length $s$.

We note that $\epsilon_{\text{A}}$,$\epsilon_{\text{B}}$ -- thereby
$\epsilon_{\text{AB}}$ in scenario (i), eq.\ (\ref{eq epsAB}), and
$\epsilon_{_{\pm}}$ in scenario (ii), eq.\ (\ref{eq epspm}) -- as well as the
setting probabilities $p_{ij}$ must be estimated in some way, presumably from
experimental characterization of the setting choice generation process. To
preserve the statistical conclusions, the used values of $\epsilon_{\text{A}}%
$, $\epsilon_{\text{B}}$, and $p_{11}$ should be conservative overestimates,
while $p_{12},p_{21},p_{22}$ should be conservative underestimates. Estimates
of this kind, including $p$-values for $\epsilon_{\text{A}}$,$\epsilon
_{\text{B}}$ have recently been reported for phase-diffusion random number
generators\ \cite{Abel2015,Mitc2015}. The $p$-value for $\epsilon_{\text{A}}%
$,$\epsilon_{\text{B}}$ can be taken into account by including the failure
probability $q_{\text{f}}$ into the process counting procedure, explained in
the previous paragraph. In general then, an experiment can thus lead to two
$p$-values, one for the process value and one for the $p_{ij}$ estimates.
These $p$-values can be used in a single test, for example using the
Bonferroni method:\ to reach significance $\alpha$, perform two separate
hypothesis tests of the two hypotheses (bounded $p_{ij}$ and local realism)
with significance $\alpha/2$.

For random number generators with small bias\ \cite{Abel2015,Mitc2015}, it
might be more efficient to quantify with $\epsilon_{\text{A}}$,$\epsilon
_{\text{B}}$ the excess predictability beyond probability $\frac{1}{2}$,
despite the presence of the bias. Expression (\ref{eq P(a)}) then becomes
$\tfrac{1}{2}\,(1-\epsilon_{\text{A}})\leq p(a|\mu)\leq\tfrac{1}%
{2}\,(1+\epsilon_{\text{A}})$, and similar for Bob. This has the advantage
that it suffices to estimate $\epsilon_{\text{A}}$,$\epsilon_{\text{B}}$ --
which in this definition now include both bias itself and excess
predictability beyond bias -- and their failure probability $q_{\text{f}}$, so
that estimates of the $p_{ij}$ are not required. The expressions
(\ref{eq Je(n)}), (\ref{eq rke}), and (\ref{eq peLR}) still hold, with all
four $p_{ij}=\tfrac{1}{4}$; in this case, an experiment leads to only one
$p$-value (for the process value). This procedure was used in
Ref.\ \cite{Gius2015}.

We finally remark that the Hoeffding bounds used above are not optimal and
better bounds are known\ \cite{Bier2015}, and that there are elegant methods
of testing local realism even without assuming any specific form of a Bell
inequality. They use the Kullback-Leibler divergence\ \cite{Kull1951}, which
measures the mathematical difference of the probability distribution obtained
from experimental data and that of any given local realist model. We refer the
reader to Refs.\ \cite{vanD2005,Zhan2010,Zhan2011,Zhan2013,Knil2015}.

\section{Conclusion}

A Bell test claiming violation of the CH/Eberhard-inequality bound by some few
standard deviations could suffer from an incomplete consideration of the task
at hand. Even disregarding world views such as superdeterminism that are
inaccessible to the scientific method, it is possible to enforce space-like
separation only up to a limit due to imperfections in even state-of-the-art
setting generators. In turn, to truly violate local realism in photonic Bell
tests, it is necessary to modify the CH/Eberhard-inequality based on the known
imperfections of the setting generator in use. We showed how to derive such
modifications in two different physical scenarios. Moreover, in the
statistical analysis we applied Doob's optional stopping theorem which
dramatically reduces the run time for reasonable experimental parameters.

\section*{Acknowledgments}

We thank R. Gill for valuable remarks on Doob's optional stopping theorem, and
W. Plick, S. Ramelow, and A. Zeilinger for detailed comments on the
manuscript. We further acknowledge helpful discussions with S. Glancy, B.
Habrich, M. Horne, A. Khrennikov, E. Knill, S. W. Nam, M. Paw{\l }owski, K.
Phelan, T. Scheidl, L. K. Shalm, R. Ursin, M. Versteegh, H. Weier, S.
Wengerowsky, B. Wittmann, and Y. Zhang. J.K.\ acknowledges support by the EU
Integrated Project SIQS. M.G.\ acknowledges support by the SFB and the CoQuS
program of the FWF (Austrian Science Fund) as well as support by the Austrian
Ministry of Science, Research and Economy through the program QUESS. M.W.M.
acknowledges support by the European Research Council project AQUMET (Grant
Agreement No.\ 280169), European Union Project QUIC (Grant Agreement
No.\ 641122), Spanish MINECO under the Severo Ochoa programme (Grant
No.\ SEV-2015-0522) and projects MAGO (Grant No.\ FIS2011-23520) and EPEC
(Grant No.\ FIS2014- 62181-EXP), Catalan AGAUR 2014 SGR Grant No.\ 1295, and
by Fundació Privada CELLEX.

\end{document}